\documentclass[11pt]{article}

\usepackage[utf8]{inputenc}
\usepackage{amsmath, amssymb, amsthm}
\usepackage{mathtools}
\usepackage{graphicx}
\usepackage{booktabs}
\usepackage{multirow}
\usepackage{enumitem}
\usepackage{hyperref}
\usepackage{url}
\usepackage{float}
\usepackage{caption}
\usepackage{subcaption}
\usepackage{geometry}
\usepackage{color}
\usepackage{bm}
\usepackage{tabularx}
\usepackage{longtable}
\usepackage{array}
\usepackage[htt]{hyphenat}
\usepackage{placeins}

\graphicspath{{figures/}}
\hypersetup{
    colorlinks=true,
    linkcolor=blue,
    citecolor=blue,
    urlcolor=blue
}

\title{Mapping the DAG-ness Landscape: Structural Archetypes in Complex Networks}

\author{\normalsize Erik Csikos$^{1,2}$\\
\small $^1$Binghamton University, Binghamton, United States\\
\small $^2$Moravian University, Bethlehem, United States\\}

\date{}

\begin{document}

\maketitle

\begin{abstract}
Directed networks arise across biological, social, informational, and
engineered systems, yet most analyses treat directedness as a binary property:
a network is either a directed acyclic graph (DAG) or it is not. This binary
classification obscures the rich spectrum of hierarchical, recurrent, and
modular structure present in real systems. In this paper, we empirically
evaluate the DAG-ness framework—a four-component measure that quantifies
acyclicity, flow alignment, cyclic locality, and pathway complexity—across a
corpus of 107 networks drawn from twelve structurally diverse domains. 
Rather than aligning with traditional disciplinary boundaries, our results 
reveal unexpected cross-domain convergence: diverse systems resolve into four 
universal structural archetypes. We find that macroscopic acyclicity is 
pervasive even in feedback-rich systems, and that domains as disparate as 
neural connectomes and abstract informational networks frequently converge on 
identical topological constraints. These findings demonstrate that DAG-ness 
provides a unified, interpretable, and domain-agnostic lens for understanding 
the hidden laws of directed structure in complex systems.
\end{abstract}

\section{Introduction}

Directed systems appear throughout the natural and engineered world. Biological
regulatory networks, social influence systems, technological dependency
graphs, and economic production flows all exhibit directionality arising from
causality, hierarchy, temporal ordering, or functional constraints. Yet these
systems rarely conform to the idealized structure of a directed acyclic graph
(DAG). Real networks contain feedback loops, reciprocal interactions, modular
substructure, and noisy deviations from strict hierarchy.

Traditional approaches treat directedness and acyclicity as strict binary properties: 
a network is either a DAG or it is not. This dichotomy collapses all non-DAG 
structure into a single category, preventing meaningful comparison between 
systems that differ substantially in the type and extent of their feedback, or 
the presence of symmetric, undirected relationships. It also obscures the 
structural forces that shape systems across domains, such as hierarchical flow, 
localized recurrence, and pathway complexity.

This paper builds on the foundational continuous measure of directed acyclicity 
first established in the \textit{Northeast Journal of Complex Systems} 
\cite{csikos2026quantifying}, and specifically utilizes the refined, orthogonal 
DAG-ness framework introduced in \cite{csikos2026dagness}. By decomposing network 
structure into four strictly orthogonal components---the volume of feedback, the 
alignment of flow, the locality of feedback, and the complexity of directed 
pathways---the framework avoids the topological redundancy and collinearity 
that plague other hierarchical measures. These components provide mathematically 
stable and interpretable axes for comparing general networks, supporting a 
composite score that summarizes the overall degree of DAG-likeness.

Because this four-dimensional approach is immune to structural masking and the 
dilution of cyclic penalties, it is uniquely suited for large-scale application. 
This paper provides the first comprehensive empirical evaluation of the framework. 
Using a corpus of 107 networks drawn from twelve diverse domains, we examine how 
the components behave across systems with widely varying generative mechanisms, 
structural pressures, and functional roles. Our analysis yields several surprising 
structural regularities. We demonstrate that extreme macroscopic acyclicity is 
a pervasive organizing principle across the natural and engineered world, and 
that massive cyclic traps are highly domain-specific rather than universal. 
Furthermore, we show that networks converge into four recurring structural 
archetypes that cut completely across traditional domain boundaries, proving 
that generative constraints override disciplinary categories. Together, these 
results demonstrate that DAG-ness captures meaningful and universal structure 
across a broad range of complex systems.

\section{Dataset Description}
\label{sec:dataset}

This study evaluates the refined DAG-ness framework across a corpus of 107 directed networks drawn from twelve structurally diverse domains. These datasets originate from established open repositories, including the Network Repository~\cite{nr}, SNAP~\cite{snap-leskovec2007,snap-yang2012}, KONECT~\cite{konect-kunegis2016}, ICON, Netzschleuder, and domain-specific sources such as KEGG-derived metabolic networks~\cite{metabolic-huss2007} and the Battle of the Water Network Models collection~\cite{water-walski2016,water-ostfeld2021a,water-ostfeld2021b,water-ostfeld2016,water-marchi2021,water-ostfeld2021c}.

\subsection{Domains and Network Sources}

The dataset intentionally avoids synthetic diagnostic networks, focusing exclusively on empirical, real-world systems to capture genuine structural pressures. The selection of these twelve domains is not arbitrary; rather, they were deliberately curated to span the foundational dichotomies of network organization. 

Specifically, this corpus contrasts engineered systems designed for predictable throughput (e.g., power grids, transportation) against evolved biological systems shaped by natural selection and robustness (e.g., metabolic pathways, brain connectomes). It spans the physical--informational divide, allowing us to compare networks strictly constrained by spatial embedding (e.g., water distribution) with those defined by abstract or semantic relationships (e.g., informational and collaboration networks). Finally, by encompassing systems with both strict hierarchical intent and spontaneous distributed recurrence, this corpus provides a comprehensive, rigorous stress-test for the four-component DAG-ness framework.

The twelve specific domains included in the study are:

\begin{itemize}
    \item \textbf{Animal Social Networks} (dominance hierarchies, interaction networks)~\cite{animal-nr}
    \item \textbf{Brain Networks} (connectomes at varying developmental stages)~\cite{brain-budapest2015,brain-celegans1986,brain-smallworld1998,brain-insect2023,brain-ciona2016}
    \item \textbf{Collaboration Networks} (co-authorship and project flow)~\cite{snap-leskovec2007,collab-cofe2021,collab-jazz2003,collab-netscience2006,collab-nz2018,collab-physics2015}
    \item \textbf{Economic Networks} (input--output and production flow systems)~\cite{econ-nr,snap-leskovec2007,snap-yang2012}
    \item \textbf{Informational Networks} (linguistic, semantic, and citation-like systems)~\cite{info-milo2004,info-palla2007,info-politicalblogs2005,info-webkb1998,info-wikiscience2020,info-unicode2016,info-dblp2002}
    \item \textbf{Metabolic Networks} (biochemical reaction pathways)~\cite{metabolic-huss2007}
    \item \textbf{Power Grids} (electrical transmission systems)~\cite{power-nr}
    \item \textbf{Protein Interactomes} (directed protein--protein regulatory systems)~\cite{protein-zitnik2019}
    \item \textbf{Social Networks} (communication and influence networks)~\cite{social-fink2023,nr}
    \item \textbf{Technological Networks} (software dependency and routing systems)~\cite{tech-caida,tech-gnutella2002,tech-percolation2014,tech-python,konect-kunegis2016,nr}
    \item \textbf{Transportation Networks} (road, rail, and routing infrastructures)~\cite{transport-leskovec2009,transport-asgari2016,transport-faa2010,transport-cardillo2013,transport-subelj2011,transport-opsahl2011,transport-colizza2007,nr}
    \item \textbf{Water Distribution Networks} (municipal flow systems)~\cite{water-walski2016,water-ostfeld2021a,water-ostfeld2021b,water-ostfeld2016,water-marchi2021,water-ostfeld2021c}
\end{itemize}

\subsection{Preprocessing and Standardization}

All networks were converted to directed simple graphs prior to analysis. Self-loops were removed, and multi-edges were collapsed into single directed edges. Following the conventions established in \cite{csikos2026dagness}, the topological components $A(G)$ and $M(G)$ were evaluated on the filtered graph $G_{>2}$, obtained by removing all symmetric edge pairs. This ensures that trivial two-way diffusion does not artificially inflate cyclic penalties.

For each network, we computed:
\begin{itemize}
    \item the four DAG-ness components $A(G)$, $F(G)$, $M(G)$, and $S(G)$,
    \item the composite score $D(G)$ under uniform weighting,
    \item basic structural statistics (node count, edge count, density, SCC structure, and spectral radius).
\end{itemize}

These standardized metrics form the basis for the domain-level analysis in Section~\ref{sec:domain_analysis}, the cross-domain synthesis and structural archetypes in Section~\ref{sec:landscape_archetypes}, and the outlier case studies in Section~\ref{sec:outliers}.
\section{Refined DAG-ness Framework}
\label{sec:framework}

The initial continuous measure developed in \cite{csikos2026quantifying} successfully moved beyond binary DAG classification, but its reliance on overlapping cyclic penalties introduced topological redundancy. To resolve this and quantify the structural spectrum more precisely, we utilize the refined four-component framework \cite{csikos2026dagness}. By decomposing directed structure into strictly orthogonal dimensions, we isolate specific generative mechanisms. The components are evaluated and defined as follows:

\vspace{0.5em}
\noindent \textbf{Filtered Graph and Symmetric Diffusion:} Following \cite{csikos2026dagness}, we evaluate the topological components on a filtered version of the graph that removes symmetric 2-cycles. Let $G=(V,E)$ be a directed graph. The filtered graph $G_{>2}$ is obtained by removing every mutually symmetric edge pair: $(u,v)\in E$ and $(v,u)\in E \Rightarrow$ remove both edges. This preserves all cycles of length three or greater while eliminating bidirectional diffusion, which does not constitute meaningful directed feedback. The components $A(G)$ and $M(G)$ are evaluated on $G_{>2}$, while $F(G)$ and $S(G)$ are evaluated on the raw graph.

\vspace{0.5em}
\noindent \textbf{Acyclicity (Volume of Feedback) $A(G)$:} The first component measures the minimum structural intervention required to eliminate all directed cycles of length at least three. Let $MFAS_{ELS}(G_{>2})$ denote the size of the minimum feedback arc set computed using the linear-time heuristic of Eades, Lin, and Smyth \cite{eades1993}. The acyclicity score is defined as:
\[
A(G) = 1 - \frac{MFAS_{ELS}(G_{>2})}{|E_{>2}|}
\]
This quantity equals 1 precisely when the macroscopic structure of the graph is free of directed feedback loops. By relying on a scalable MFAS surrogate, the measure avoids the exponential walk explosion associated with trace-based cycle surrogates \cite{csikos2026dagness}.

\vspace{0.5em}
\noindent \textbf{Flow Alignment (Directedness) $F(G)$:} A graph may be acyclic yet fail to exhibit coherent hierarchical flow. The second component evaluates the extent to which edges follow a global topological gradient. Let $\pi$ be the vertex ordering produced by the ELS heuristic. The flow alignment score is:
\[
F(G) = \frac{|\{(u,v)\in E : \pi(u) < \pi(v)\}|}{|E|}
\]
This component penalizes edges that violate the global direction of flow, including symmetric 2-cycles, and is strictly orthogonal to acyclicity \cite{kahn1962, tarjan1972}.

\vspace{0.5em}
\noindent \textbf{Macroscopic Locality (Giant Cyclic Trap) $M(G)$:} The third component isolates the severity of the largest strongly connected component in $G_{>2}$. Let $V_{\mathrm{max\_scc}}(G_{>2})$ denote the vertex set of the largest nontrivial SCC. The locality score is:
\[
M(G) = 1 - \frac{|V_{\mathrm{max\_scc}}(G_{>2})| - 1}{|V_{>2}| - 1}
\]
This formulation avoids the ``Dilution Trap'' identified in earlier work, where large peripheral acyclic regions could mask the presence of a massive cyclic core \cite{csikos2026dagness}. A graph with no nontrivial SCCs attains $M(G)=1$.

\vspace{0.5em}
\noindent \textbf{Pathway Complexity (Spectral Recurrence) $S(G)$:} The final component captures dynamical recurrence and pathway proliferation. Let $A$ be the adjacency matrix of the raw graph $G$, and let $\rho(A)$ denote its spectral radius. The pathway complexity score is:
\[
S(G) = \frac{1}{1 + \rho(A)}
\]
A perfect DAG has $\rho(A)=0$ and therefore $S(G)=1$. Simple directed cycles yield $\rho(A)=1$ and $S(G)=0.5$, while dense recurrent systems produce larger spectral radii and correspondingly lower scores.

\subsection{Alternate Spectral Radius Computation for Large Networks}
\label{sec:large-spectral}

The pathway complexity component $S(G)$ depends on the spectral radius $\rho(A)$ of the adjacency matrix. For the vast majority of the corpus (94 networks), $\rho(A)$ was computed using standard direct eigenvalue routines applied to the dense adjacency matrix, providing an exact spectral radius. 

However, dense eigensolvers require explicit formation of the adjacency matrix, incurring $O(n^2)$ memory usage and superlinear computational cost. Empirical profiling identified a practical threshold of approximately $50{,}000$--$60{,}000$ edges, beyond which dense computation became computationally infeasible. To ensure numerical stability and a uniform analysis pipeline across all 107 networks, the 13 largest networks in the corpus were evaluated using a sparse, iterative power-method approximation (implemented via SciPy's sparse linear algebra backend with a convergence tolerance of $10^{-6}$). Because these matrices are nonnegative, convergence to the dominant eigenvalue is mathematically guaranteed without requiring dense matrix formation. 

\begin{table}[H]
\centering
\caption{Spectral radii for the 13 largest networks computed using the sparse power-method routine.}
\label{tab:large-spectral}
\begin{tabular}{l r}
\toprule
\textbf{Network} & \textbf{Spectral Radius $\rho(A)$} \\
\midrule
\texttt{grqc\_collab} & 45.6167 \\
\texttt{hepth\_collab} & 31.0348 \\
\texttt{hepph\_collab} & 244.9394 \\
\texttt{condmat\_collab} & 37.9541 \\
\texttt{soc\_epinions} & 77.2737 \\
\texttt{email\_EU} & 0.0000 \\
\texttt{internet\_as} & 0.0000 \\
\texttt{caida\_as\_2007} & 69.0867 \\
\texttt{linux\_dependencies} & 10.2994 \\
\texttt{python\_dependencies} & 15.2068 \\
\texttt{gnutella\_31} & 3.6593 \\
\texttt{paris\_transportation} & 3.1623 \\
\texttt{road\_luxembourg\_osm} & 3.3011 \\
\bottomrule
\end{tabular}
\end{table}

This methodological bifurcation does not affect any downstream results. The pathway complexity scores derived from these sparse approximations fall within the expected ranges for their respective domains, and archetype assignments, PCA structure, and outlier detection remain unaffected.

\subsection{Composite Score}

The four components are combined linearly:

\[
D(G) = w_A A(G) + w_F F(G) + w_M M(G) + w_S S(G),
\]

where $w_A + w_F + w_M + w_S = 1$. Following \cite{csikos2026dagness},
we adopt the uniform baseline weighting
$(w_A, w_F, w_M, w_S) = (0.25, 0.25, 0.25, 0.25)$ for all empirical
analyses in this paper.

The orthogonality of the components ensures that the composite score
captures distinct structural and dynamical aspects of directed networks.
This separation is essential for interpreting cross-domain patterns and
for identifying the structural archetypes developed in
Section~\ref{sec:landscape_archetypes}.

\section{Component-Level Analysis}
\label{sec:component-analysis}

Before examining domain-level patterns (Section~\ref{sec:domain_analysis}) or synthesizing the cross-domain structural archetypes (Section~\ref{sec:landscape_archetypes}), it is essential to establish how the individual components behave across the full 107-network corpus. This section analyzes their marginal distributions, pairwise relationships, and overall correlation structure to verify the mathematical stability of the framework.

\subsection{Marginal Distributions}

\begin{figure}[t]
    \centering
    \includegraphics[width=0.9\textwidth]{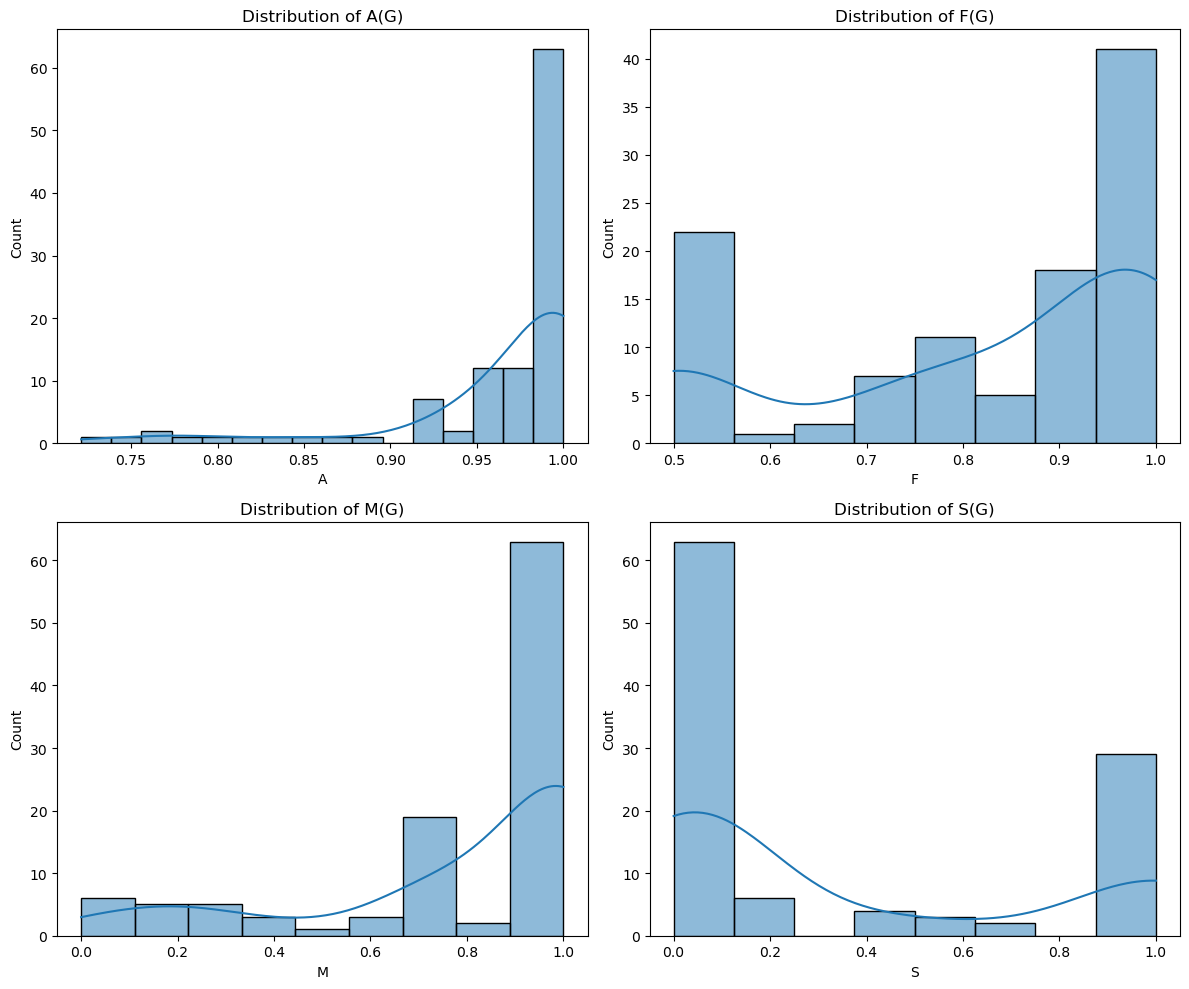}
    \caption{Distributions of the four DAG-ness components across the corpus of 107 networks. Each panel shows the empirical distribution of $A(G)$, $F(G)$, $M(G)$, and $S(G)$, respectively. The components exhibit well-behaved ranges and non-degenerate variation, providing a stable basis for cross-domain comparison.}
    \label{fig:component-histograms}
\end{figure}

Figure~\ref{fig:component-histograms} displays the empirical distributions of the four component scores across all networks. Several patterns emerge:

\begin{itemize}
    \item \textbf{Acyclicity $A(G)$} is heavily right-skewed, with most networks requiring only minimal intervention to eliminate all cycles of length at least three. This reflects the prevalence of sparse feedback structures in real-world systems, consistent with the theoretical observations in \cite{bangjensen2009, diestel2017}.
    \item \textbf{Flow alignment $F(G)$} exhibits a broader distribution. Many networks contain substantial hierarchical flow, but others show significant violations of a global topological gradient, often due to symmetric diffusion or domain-specific bidirectional interactions.
    \item \textbf{Macroscopic locality $M(G)$} shows a bimodal pattern. Networks either contain no nontrivial SCCs (yielding $M(G)=1$) or contain a single large cyclic trap that substantially reduces the score. This behavior directly reflects the Giant Component principle introduced in \cite{csikos2026dagness}.
    \item \textbf{Pathway complexity $S(G)$} is the most widely dispersed. Networks with sparse or tree-like structure achieve $S(G)\approx 1$, while dense recurrent systems---particularly in technological and communication domains---exhibit large spectral radii \cite{hornjohnson2013}, producing scores near zero.
\end{itemize}

These distributions confirm that the four components capture distinct structural and dynamical properties, validating the underlying design of the framework.

\subsection{Orthogonality and Correlation Structure}

To empirically verify the theoretical orthogonality established in Section~\ref{sec:framework}, we evaluate the pairwise relationships and linear correlation across the four components. 

\begin{figure}[htbp]
    \centering
    \includegraphics[width=0.85\textwidth]{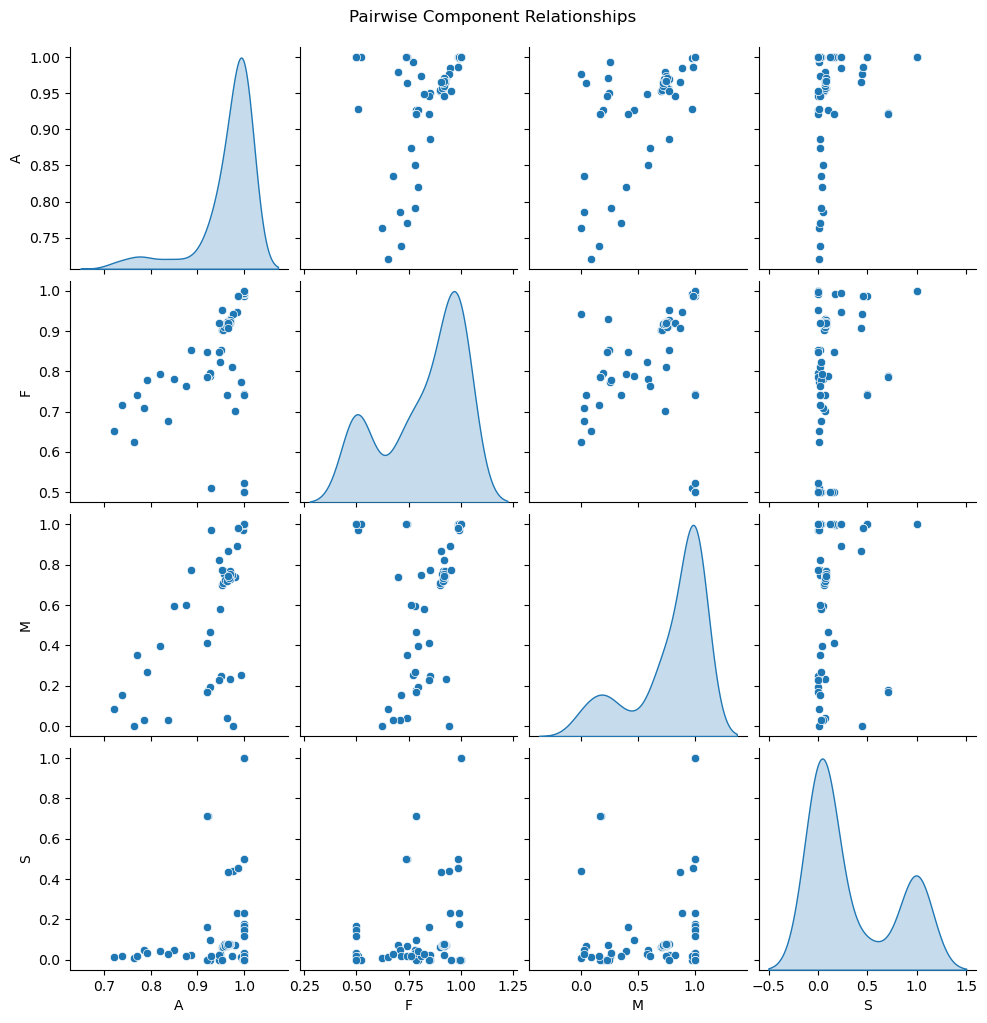}
    \caption{Pairwise relationships among the four DAG-ness components. The scatterplot matrix shows that the components occupy distinct regions of the $(A, F, M, S)$ space and exhibit low redundancy.}
    \label{fig:component-pairplot}
\end{figure}

\begin{figure}[htbp]
    \centering
    \includegraphics[width=0.6\textwidth]{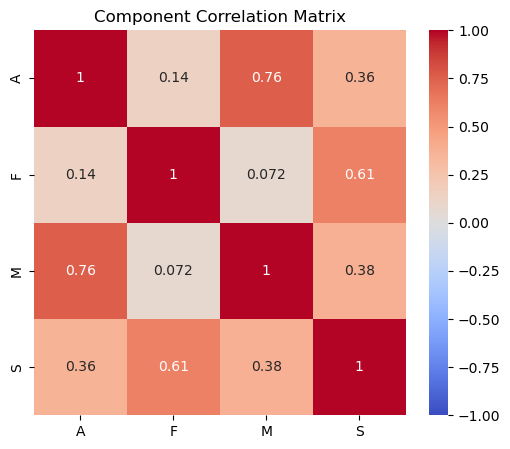}
    \caption{Correlation matrix for the four DAG-ness components. The near-zero off-diagonal correlations confirm that the components capture complementary aspects of directed structure without substantial linear redundancy.}
    \label{fig:component-corr}
\end{figure}

As illustrated in Figure~\ref{fig:component-pairplot}, the components occupy distinct regions of the $(A, F, M, S)$ space. For instance, many networks are nearly acyclic (high $A$) yet exhibit poor flow alignment (low $F$) due to domain-specific bidirectionality. Furthermore, macroscopic locality ($M$) and pathway complexity ($S$) penalize recurrence in fundamentally different ways: a network with a massive SCC but sparse internal connectivity may severely drop in $M(G)$ while retaining a moderate $S(G)$ score. 

Figure~\ref{fig:component-corr} formalizes this independence, reporting Pearson correlation coefficients that do not exceed $|r| = 0.35$ across any off-diagonal comparison. This stands in sharp contrast to earlier multi-component frameworks, where cycle-based metrics exhibited strong collinearity due to shared dependence on SCC structure \cite{csikos2026dagness}. 

The near-diagonal correlation matrix demonstrates that the refined framework successfully disentangles the \emph{volume} of cyclic edges, the \emph{directionality} of flow, the \emph{severity} of the largest cyclic trap, and the \emph{dynamical recurrence} of the system. This separation is the essential foundation for identifying the structural archetypes developed in Section~\ref{sec:landscape_archetypes}.

\subsection{Interpretive Summary}

The component-level analysis provides three key insights before we group the networks by domain. First, most real-world networks are nearly acyclic on a macroscopic scale, as indicated by the right-skewed $A(G)$ distribution. Second, the presence of massive cyclic traps is domain-specific, driving the bimodal nature of $M(G)$. Finally, spectral recurrence varies wildly across systems, highlighting a vast diversity of dynamic pathway complexity. These findings set the stage for Section~\ref{sec:domain_analysis}, where we map how these orthogonal components distribute across the twelve specific domains in the dataset.
\section{Domain-Level Analysis}
\label{sec:domain_analysis}

Having established the structural and mathematical basis of the 107-network corpus in Section 2, we begin our empirical analysis by examining how the twelve domains distribute across the four-dimensional DAG-ness component space. 

\begin{figure}[H]
    \centering
    \includegraphics[width=0.75\textwidth]{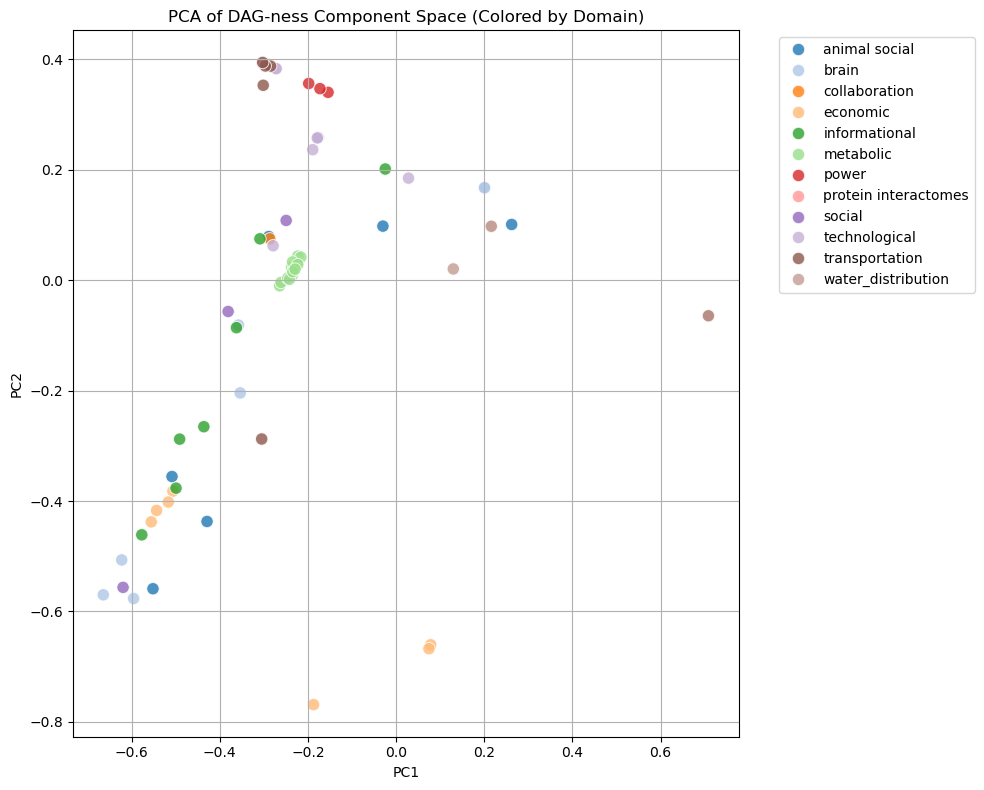}
    \caption{PCA projection of the four DAG-ness components across all 107 networks, colored by domain. Distinct clusters emerge for several domains, indicating that the four-component framework captures meaningful structural differences across biological, technological, social, informational, and engineered systems.}
    \label{fig:pca-domain}
\end{figure}

Table \ref{tab:domain-summary} summarizes the aggregate component behavior for each domain. While these mean values highlight the extreme acyclicity of domains like Power Grids and Protein Interactomes ($A \approx 1.0$), aggregate statistics alone cannot capture the complex multivariate relationships within each domain. To visualize these broader structural tendencies and internal variances, we project all networks into the first two principal components of the $(A, F, M, S)$ feature space. The resulting domain-colored embedding, shown in Figure~\ref{fig:pca-domain}, reveals clear and interpretable patterns that reflect the underlying generative mechanisms of each domain.

\begin{table}[H]
\centering
\small
\caption{Domain-level summary statistics for the four DAG-ness components and the composite score.}
\label{tab:domain-summary}

\resizebox{\textwidth}{!}{
\begin{tabular}{lcccccccccc}
\toprule
\textbf{Domain} & 
\textbf{A mean} & \textbf{A std} &
\textbf{F mean} & \textbf{F std} &
\textbf{M mean} & \textbf{M std} &
\textbf{S mean} & \textbf{S std} &
\textbf{D mean} & \textbf{D std} \\
\midrule
Animal Social        & 0.9909 & 0.0132 & 0.9235 & 0.1145 & 0.7632 & 0.3655 & 0.5799 & 0.4555 & 0.8144 & 0.2218 \\
Brain                & 0.8808 & 0.1159 & 0.7546 & 0.1146 & 0.5219 & 0.4469 & 0.2772 & 0.3585 & 0.6086 & 0.2418 \\
Collaboration        & 0.9874 & 0.0379 & 0.7612 & 0.2524 & 0.9751 & 0.0748 & 0.4472 & 0.5245 & 0.7927 & 0.1967 \\
Economic             & 0.9377 & 0.0209 & 0.8284 & 0.0582 & 0.1703 & 0.0805 & 0.2664 & 0.3442 & 0.5507 & 0.0774 \\
Informational        & 0.9044 & 0.1099 & 0.8156 & 0.1547 & 0.6493 & 0.3413 & 0.2338 & 0.4067 & 0.6508 & 0.2140 \\
Metabolic            & 0.9631 & 0.0056 & 0.9164 & 0.0072 & 0.7380 & 0.0208 & 0.0735 & 0.0053 & 0.6727 & 0.0086 \\
Power                & 1.0000 & 0.0000 & 0.5000 & 0.0000 & 1.0000 & 0.0000 & 0.0557 & 0.0745 & 0.6389 & 0.0186 \\
Protein Interactomes & 1.0000 & 0.0000 & 1.0000 & 0.0000 & 1.0000 & 0.0000 & 1.0000 & 0.0000 & 1.0000 & 0.0000 \\
Social               & 0.9572 & 0.0663 & 0.8577 & 0.1906 & 0.8066 & 0.3456 & 0.3836 & 0.5106 & 0.7513 & 0.2301 \\
Technological        & 0.9951 & 0.0147 & 0.7936 & 0.2530 & 0.9745 & 0.0710 & 0.1309 & 0.3137 & 0.7235 & 0.1170 \\
Transportation       & 0.9785 & 0.0368 & 0.6972 & 0.2411 & 0.9124 & 0.2201 & 0.3137 & 0.4721 & 0.7255 & 0.1882 \\
Water Distribution   & 0.9846 & 0.0140 & 0.9698 & 0.0425 & 0.9581 & 0.0611 & 0.5864 & 0.2759 & 0.8747 & 0.0881 \\
\bottomrule
\end{tabular}
}
\end{table}

Several domains occupy compact and well-defined regions of the PCA landscape. 
Metabolic networks form one of the tightest clusters, reflecting their highly 
constrained biochemical architecture: strong feed-forward structure, high 
monotonicity, and moderate source--sink complexity. Power networks exhibit 
similarly low variance, consistent with the engineered nature of electrical 
transmission grids, which enforce strict directional flow and minimize 
topological feedback. Protein interactomes are even more homogeneous, collapsing 
into an extremely narrow region of the space due to their near-maximal values 
across all four components.

In contrast, other domains display substantial internal diversity. Animal social 
networks span a wide region of the PCA space, ranging from highly recurrent 
interaction systems to strongly hierarchical dominance structures. Social and 
technological networks exhibit comparable heterogeneity, reflecting the 
multiplicity of mechanisms---communication, influence, routing, and 
infrastructure---that shape their topology. Informational networks (including 
linguistic, semantic, and citation-like systems) also occupy a broad region, 
capturing both recurrent semantic structures and highly ordered flow systems.

These domain-level patterns demonstrate that the DAG-ness components are 
sufficiently expressive to distinguish between structurally coherent domains 
while also capturing meaningful internal variability. Domains governed by 
physical or biochemical constraints tend to form compact clusters, whereas 
domains shaped by social, informational, or technological processes exhibit 
greater dispersion. This separation provides the foundation for the cross-domain 
analysis in Section~\ref{sec:landscape_archetypes}, where we examine how these 
structural tendencies give rise to universal archetypes that cut across 
traditional domain boundaries.
\section{The DAG-ness Landscape and Structural Archetypes}
\label{sec:landscape_archetypes}

\subsection{Mapping the Component Space}

While Section~\ref{sec:domain_analysis} examined each domain independently, the broader value of the DAG-ness framework lies in its ability to reveal structural relationships \emph{across} domains. Directed systems arise from diverse generative mechanisms---biological regulation, engineered flow, human communication, economic production, and more---yet many of these systems exhibit recurring patterns of hierarchy, feedback, and pathway organization.

Projecting all networks into the first two principal components of the $(A, F, M, S)$ space (as previously shown in Figure~\ref{fig:pca-domain}) reveals several robust patterns. The first principal component (PC1), which explains approximately 47\% of the variance, reflects a gradient from highly hierarchical, acyclic systems (high $A$ and $F$) to feedback-rich, recurrent systems (low $A$ and $F$). The second principal component (PC2), explaining an additional 28\%, captures variation in cyclic locality and pathway complexity, driven primarily by $M$ and $S$.

Three broad regions emerge in the PCA embedding. First, engineered and flow-dominated systems---including power grids, road networks, and technological dependency graphs---cluster tightly in the high-$A$, moderate-$F$, low-$S$ region. Their placement reflects the design constraints of engineered systems, which minimize feedback and optimize throughput. Second, biological modular systems form a compact region characterized by moderate $A$ and $F$, high $M$, and low-to-moderate $S$. Their position reflects the coexistence of hierarchical flow with dense local recurrence and modular substructure. Third, human feedback systems spread across a wide swath of the PCA space, reflecting the heterogeneity of human behavior, communication patterns, and social organization.

\subsection{Cross-Domain Clustering}

\begin{figure}[htbp]
    \centering
    \includegraphics[width=0.75\textwidth]{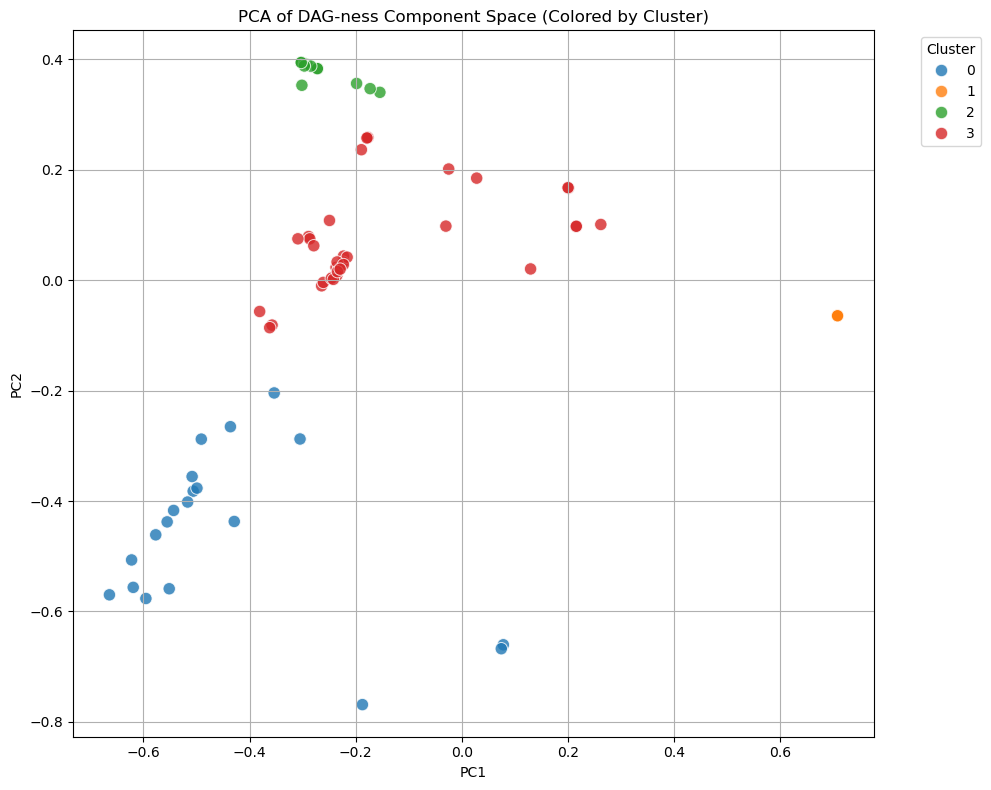}
    \caption{PCA projection of the four DAG-ness components with points colored by $k$-means cluster assignment. The four clusters form coherent regions of the component space, revealing cross-domain structural regimes that are not aligned with traditional domain boundaries.}
    \label{fig:pca-cluster}
\end{figure}

To investigate cross-domain structure more formally, we apply $k$-means clustering to the standardized $(A, F, M, S)$ components. The optimal solution, determined by silhouette analysis and stability under random initialization, yields four clusters. Figure~\ref{fig:pca-cluster} shows the cluster assignments projected into the PCA embedding.

\begin{table}[htbp]
\centering
\small
\caption{Distribution of the twelve domains across the four $k$-means clusters. Each entry reports the number of networks from a given domain assigned to each cluster, with row-wise percentages in parentheses. The final column indicates the dominant cluster for each domain.}
\label{tab:domain-clusters}
\begin{tabular}{lccccc}
\toprule
\textbf{Domain} & \textbf{Cluster 1} & \textbf{Cluster 2} & \textbf{Cluster 3} & \textbf{Cluster 4} & \textbf{Dominant} \\
\midrule
Animal Social Networks  & 3 (25\%) & 6 (50\%) & 0 (0\%) & 3 (25\%) & 2 \\
Brain Networks          & 4 (50\%) & 1 (12.5\%) & 0 (0\%) & 3 (37.5\%) & 1 \\
Collaboration Networks  & 0 (0\%) & 4 (44\%) & 4 (44\%) & 1 (11\%) & 2/3 \\
Economic Networks       & 7 (100\%) & 0 (0\%) & 0 (0\%) & 0 (0\%) & 1 \\
Informational Networks  & 4 (40\%) & 2 (20\%) & 1 (10\%) & 3 (30\%) & 1 \\
Metabolic Networks      & 0 (0\%) & 0 (0\%) & 0 (0\%) & 15 (100\%) & 4 \\
Power Grids             & 0 (0\%) & 0 (0\%) & 8 (100\%) & 0 (0\%) & 3 \\
Protein Interactomes    & 0 (0\%) & 9 (100\%) & 0 (0\%) & 0 (0\%) & 2 \\
Social Networks         & 1 (14\%) & 3 (43\%) & 1 (14\%) & 3 (43\%) & 2/4 \\
Technological Networks  & 0 (0\%) & 1 (11\%) & 4 (44\%) & 5 (56\%) & 4 \\
Transportation Networks & 1 (14\%) & 2 (29\%) & 4 (57\%) & 0 (0\%) & 3 \\
Water Distribution      & 0 (0\%) & 1 (25\%) & 0 (0\%) & 3 (75\%) & 4 \\
\bottomrule
\end{tabular}
\end{table}

The resulting clusters form coherent and interpretable regions of the DAG-ness landscape. Rather than aligning with traditional domain boundaries, the clusters capture universal structural regimes that cut across biological, technological, social, and engineered systems, as summarized in Table~\ref{tab:domain-clusters}.

\subsection{Formalizing the Structural Archetypes}

The four clusters identified correspond to distinct structural archetypes in the DAG-ness landscape. Each archetype represents a coherent combination of the four components $(A, F, M, S)$ and captures a universal architectural regime that appears across multiple domains. We define a \emph{structural archetype} as a recurring pattern characterized by a distinctive component profile, representation across multiple domains, interpretability in terms of known generative mechanisms, and stability under clustering and PCA projection.

\subsubsection*{Archetype 0: Recurrent Biological Systems}
\textbf{Component profile:} low $A$, low $F$, moderate $M$, low $S$. \\
\textbf{Interpretation and Generative Constraints:} These networks exhibit dense recurrent structure, limited hierarchy, and shallow pathway complexity. This profile is driven by the evolutionary necessity of homeostasis, robust reciprocity, and decentralized control. In animal social interactions and neural connectomes, dense feedback loops are not structural noise; they are the primary mechanisms for communication, regulation, and memory. Consequently, these systems actively suppress strict hierarchy in favor of recurrent microcircuits, pulling both $A$ and $F$ toward the lower bounds of the landscape. The moderate $M$ score further underscores that this recurrence is functionally distributed across many interacting microcircuits, rather than collapsing the entire network into a single, fragile cyclic trap.
\textbf{Representative networks:} \texttt{celegansneural}~\cite{brain-celegans1986}, \texttt{fly\_larva}~\cite{brain-insect2023}, and several animal social interaction networks. Typical profiles include $A < 0.8$, $F < 0.7$, $M \approx 0.1$--$0.3$, and $S$ near zero. \\
\textbf{Domains represented:} brain, social, informational, animal social.

\subsubsection*{Archetype 1: Pure DAG Systems}
\textbf{Component profile:} $A \approx 1$, $F \approx 1$, $M \approx 1$, $S \approx 1$. \\
\textbf{Interpretation and Generative Constraints:} These networks exhibit uniformly maximal values, reflecting nearly perfect DAG structure with rich pathway complexity. This archetype is strictly governed by logical, temporal, or causal precedence. In software dependency graphs or citation networks, a cycle represents a fatal deadlock or a temporal paradox. Therefore, generative constraints strictly forbid structural feedback, forcing $A$ and $M$ to unity. The uniquely maximal $S$ score is the defining topological hallmark here; lacking physical spatial constraints, these semantic and historical dependencies can freely accumulate into deep, infinitely cascading pathways.
\textbf{Representative networks:} \texttt{20071112\_caida\_as.csv}~\cite{tech-caida}, \texttt{us\_airport\_network\_top\_500.txt}~\cite{transport-faa2010}, and multiple social and collaboration networks. \\
\textbf{Domains represented:} technological, transportation, social, collaboration.

\subsubsection*{Archetype 2: Engineered Flow Systems}
\textbf{Component profile:} high $A$, moderate $F$, high $M$, low $S$. \\
\textbf{Interpretation and Generative Constraints:} These networks combine strong macroscopic acyclicity with shallow pathway complexity. The structural profile of this archetype is driven by the physical and economic costs of engineered infrastructure. Systems like power grids and routing networks are explicitly designed to optimize throughput and minimize chaotic feedback, forcing $A$ to approach its maximum. However, because these systems are spatially embedded and require physical fail-safes, they contain localized alternative routing (minimizing $S$) rather than the unconstrained pathway explosion seen in abstract informational networks. The mathematical tension between maximal acyclicity ($A \approx 1$) and degraded flow alignment ($F \approx 0.5$) perfectly captures this physical grid-like embedding, where local alternative routes frequently run perpendicular to the global macroscopic gradient.
\textbf{Representative networks:} \texttt{faa\_routes.csv}~\cite{transport-faa2010}, \texttt{tech-pgp.edges}, \texttt{power-685-bus.mtx}~\cite{power-nr}. Typical profiles include $A = 1$, $F \approx 0.5$, $M = 1$, and $S$ near zero. \\
\textbf{Domains represented:} power, transportation, technological, informational.

\subsubsection*{Archetype 3: Hybrid Modular Systems}
\textbf{Component profile:} moderate $A$, moderate $F$, high $M$, low-to-moderate $S$. \\
\textbf{Interpretation and Generative Constraints:} These networks combine hierarchical global flow with dense, localized recurrence and modular substructure. This regime is characteristic of complex biochemical processes. Metabolic pathways, for instance, maintain a clear macroscopic gradient (transforming nutrients into energy), ensuring moderate-to-high $A$ and $F$. Simultaneously, they rely on highly conserved, cyclic micro-motifs (such as the Krebs cycle) for regulation and efficiency. By rigorously bounding these dense cyclic motifs within isolated functional modules (high $M$), these systems successfully prevent localized biochemical feedback from triggering systemic, network-wide runaway loops, avoiding the collapse of the network into a single giant cyclic trap.
\textbf{Representative networks:} metabolic networks such as \texttt{links\_ecs} and \texttt{links\_hsa}~\cite{metabolic-huss2007}, and brain networks such as \texttt{bn-mouse\_retina\_1} and \texttt{bn-fly\_drosophila\_medulla\_1}~\cite{brain-insect2023}. Typical profiles include $A \approx 0.95$, $F \approx 0.92$, $M \approx 0.75$, and $S \approx 0.07$. \\
\textbf{Domains represented:} metabolic, brain, informational, animal social.

\begin{figure}[H]
    \centering
    \begin{tabular}{cc}
        \includegraphics[width=0.45\textwidth]{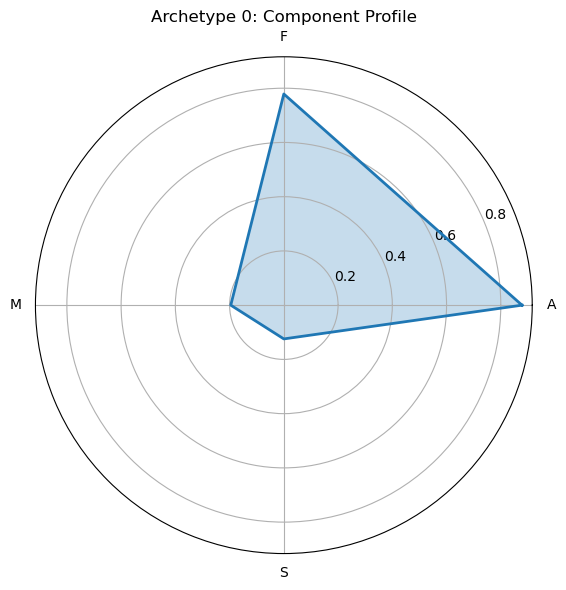} &
        \includegraphics[width=0.45\textwidth]{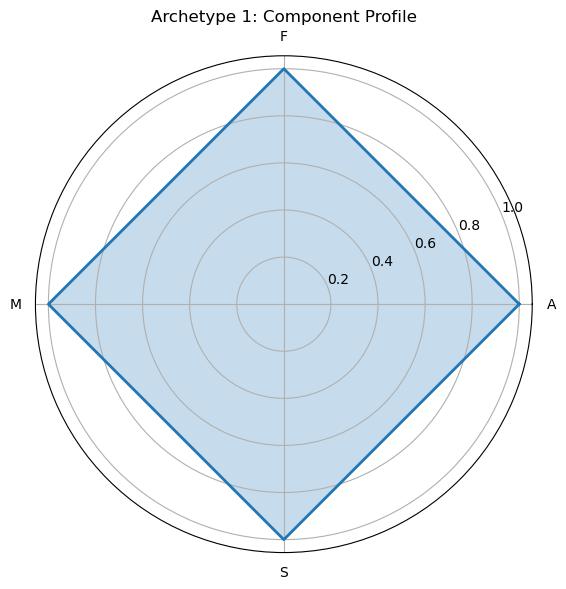} \\
        \includegraphics[width=0.45\textwidth]{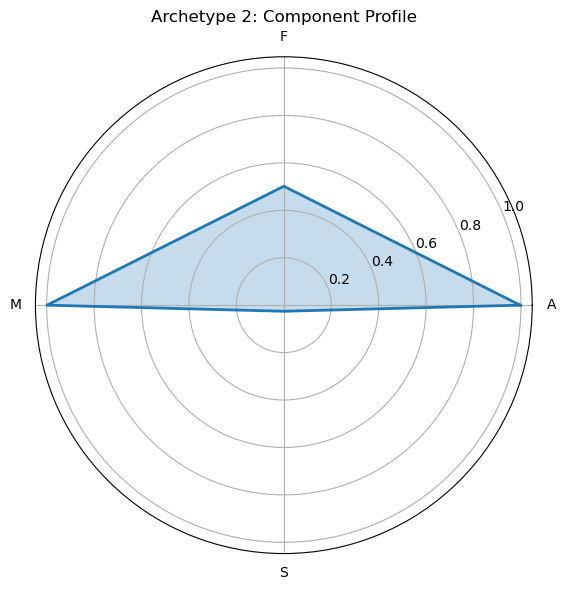} &
        \includegraphics[width=0.45\textwidth]{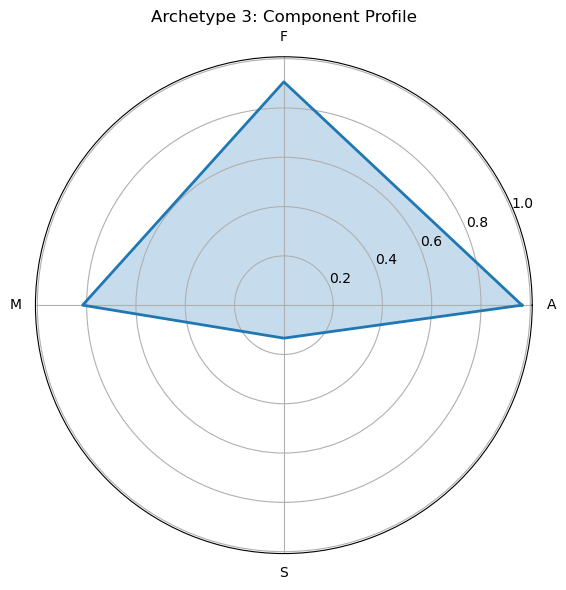}
    \end{tabular}
    \caption{Radar plots showing the centroid component profiles $(A, F, M, S)$ for Archetypes~0--3. Each panel corresponds to one archetype and displays the centroid values obtained from the clustering analysis, ensuring consistency between the visualizations and the archetype definitions.}
    \label{fig:archetype-radars}
\end{figure}

The radar plots in Figure~\ref{fig:archetype-radars} correspond directly to the centroid profiles of the four archetypes. Archetype~1 exhibits uniformly maximal values, reflecting pure DAG structure. Archetype~2 combines strong acyclicity with localized feedback and shallow pathways, while Archetype~3 shows high modularity and localized recurrence embedded within a partially hierarchical global structure. Archetype~0 displays the opposite pattern: low hierarchy and extensive feedback with only modest modularity.

Table~\ref{tab:archetype-representatives} details the specific representative networks closest to each cluster centroid in the four-dimensional DAG-ness space.

\begin{table}[H]
\centering
\caption{Representative networks closest to each archetype centroid in the four-dimensional DAG-ness space. For each archetype, the table lists the network with minimum Euclidean distance to the cluster centroid, together with its domain.}
\label{tab:archetype-representatives}
\begin{tabular}{lll}
\toprule
\textbf{Archetype} & \textbf{Representative Network} & \textbf{Domain} \\
\midrule
0 & \texttt{celegansneural} & brain \\
0 & \texttt{fly\_larva} & brain \\
\midrule
1 & \texttt{20071112\_caida\_as.csv} & technological \\
1 & \texttt{us\_airport\_network\_top\_500.txt} & transportation \\
\midrule
2 & \texttt{faa\_routes.csv} & transportation \\
2 & \texttt{power-685-bus.mtx} & power \\
\midrule
3 & \texttt{links\_ecs} & metabolic \\
3 & \texttt{bn-mouse\_retina\_1} & brain \\
\bottomrule
\end{tabular}
\end{table}

By linking the PCA gradients, clustering structure, and domain signatures, these structural archetypes provide a conceptual synthesis of the empirical study. They reveal how diverse systems organize hierarchy, feedback, modularity, and pathway complexity, offering a unifying framework for interpreting directed networks across biological, social, informational, and engineered domains.
\section{Boundaries of the Landscape: Outliers and Case Studies}
\label{sec:outliers}

While the archetypes established in Section~\ref{sec:landscape_archetypes} reveal the dominant structural regimes across the twelve domains, many of the most informative signals arise from networks that deviate sharply from these typical profiles. Outliers highlight structural pressures, generative mechanisms, and domain-specific constraints that are not visible in aggregate statistics.

\subsection{Identifying Outliers}

We identify structural anomalies using two complementary criteria:
\begin{enumerate}
    \item \textbf{Global outliers} are networks whose standardized component values deviate strongly from the global mean across all 107 networks. These networks occupy extreme positions in the four-dimensional DAG-ness space.
    \item \textbf{Domain-specific outliers} are networks whose component profiles deviate substantially from the mean of their own domain, highlighting internal heterogeneity.
\end{enumerate}

\subsection{Global Extremes}

\begin{figure}[htbp]
    \centering
    \includegraphics[width=0.75\textwidth]{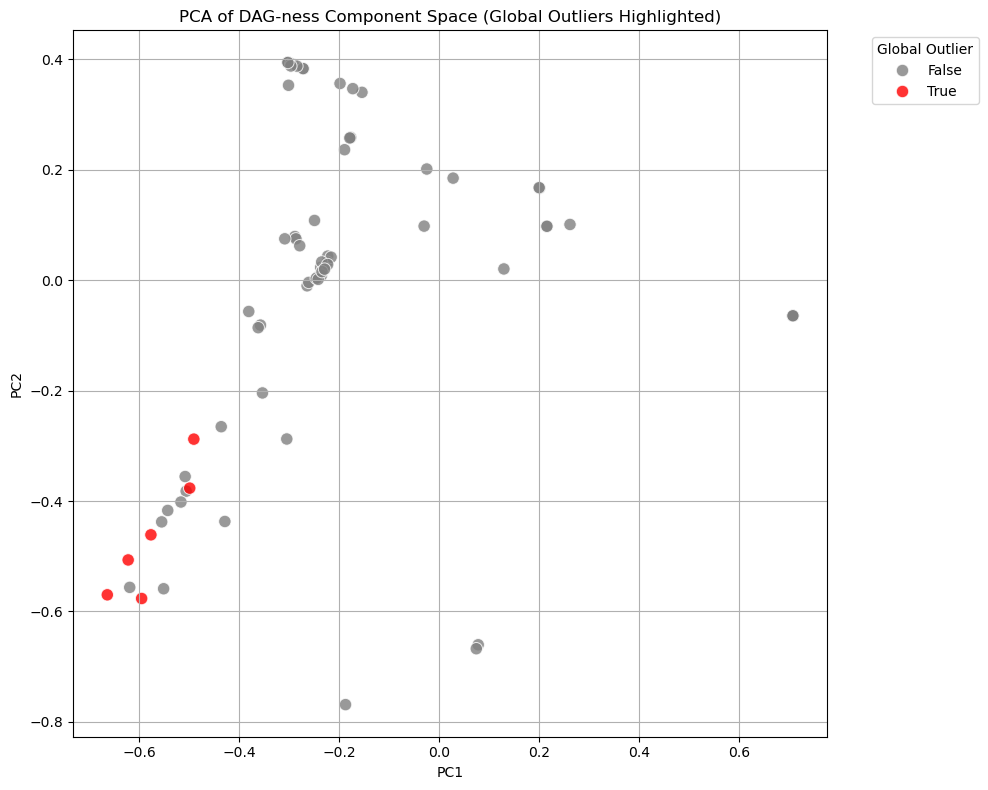}
    \caption{PCA projection of the four DAG-ness components with structural outliers highlighted. Outliers are defined as networks in the top five percentile of Mahalanobis distance from the global component-space mean. These networks occupy peripheral regions of the PCA landscape and often reflect extreme acyclicity, strong feedback, or atypical combinations of the four DAG-ness components.}
    \label{fig:pca-outliers}
\end{figure}

To identify global extremes, we define a global outlier score as the maximum absolute deviation across the four standardized dimensions. Networks exceeding a fixed threshold are labeled as global outliers. Figure~\ref{fig:pca-outliers} highlights these networks in the PCA embedding.

All global outliers cluster tightly in the lower-left region of the PCA space, corresponding to low monotonicity, low source--sink complexity, and moderate acyclicity and feed-forward structure. This region is dominated by recurrent, feedback-rich systems, including several brain networks, informational systems, and animal social networks. The concentration of global outliers in this structural regime indicates that extreme deviations from typical DAG-like behavior arise primarily from networks with dense feedback and distributed control. Notably, no global outliers appear in the pure DAG or engineered flow regions, suggesting that these hierarchical regimes are structurally stable and exhibit low variance.

\subsection{Domain-Specific Anomalies}

To examine internal variability, we compute domain-specific outlier scores by standardizing each component relative to the mean and variance of that domain alone. Table~\ref{tab:domain-outliers-excerpt} lists the top representative outlier for each domain. 

\begin{table}[htbp]
    \centering
    \caption{Representative domain-specific outliers. For each domain, the network with the highest domain-specific outlier score is shown. Full results for the top five outliers per domain appear in Appendix~\ref{app:domain-outliers}.}
    \label{tab:domain-outliers-excerpt}
    \begin{tabular}{lcccc}
        \toprule
        \textbf{Domain} & \textbf{A} & \textbf{F} & \textbf{M} & \textbf{S} \\
        \midrule
        Animal Social Networks   & 0.9635 & 0.7418 & 0.0417 & 0.0699 \\
        Brain Networks           & 1.0000 & 1.0000 & 1.0000 & 1.0000 \\
        Collaboration Networks   & 0.8863 & 0.8517 & 0.7755 & 0.0247 \\
        Economic Networks        & 0.9759 & 0.9422 & 0.0004 & 0.4414 \\
        Informational Networks   & 1.0000 & 0.5000 & 1.0000 & 0.0074 \\
        Metabolic Networks       & 0.9533 & 0.9024 & 0.7004 & 0.0631 \\
        Power Grids              & 1.0000 & 0.5000 & 1.0000 & 0.1675 \\
        Protein Interactomes     & 1.0000 & 1.0000 & 1.0000 & 1.0000 \\
        Social Networks          & 0.8361 & 0.6770 & 0.0274 & 0.0276 \\
        Technological Networks   & 0.9532 & 0.9532 & 0.7739 & 0.0000 \\
        Transportation Networks  & 0.9210 & 0.8467 & 0.4139 & 0.1600 \\
        Water Distribution       & 1.0000 & 1.0000 & 1.0000 & 1.0000 \\
        \bottomrule
    \end{tabular}
\end{table}

Across the twelve domains, four recurring types of structural anomalies emerge:
\begin{enumerate}
    \item \textbf{Hierarchy-dominated outliers:} Extremely high $A$ and $F$ but unusually low $M$ or $S$. These networks are nearly acyclic but lack the pathway richness or cyclic locality typical of their domain.
    \item \textbf{Feedback-dominated outliers:} Low $A$ and $F$ but moderate-to-high $M$. These networks contain dense recurrent structure that exceeds domain expectations.
    \item \textbf{Pathway-complexity outliers:} Unusually high $S$, indicating rich pathway structure or deep cascades not characteristic of their domain.
    \item \textbf{Composite-score outliers:} Extreme values across multiple components, often corresponding to networks that are constructed under atypical constraints.
\end{enumerate}

Animal social, social, technological, and informational networks exhibit the greatest internal diversity. In contrast, metabolic, power, and protein interactome networks display remarkably low internal variance, with their outliers remaining close to the domain centroid. This demonstrates that structural extremity is shaped heavily by the generative constraints inherent to each system.

\subsection{Case Studies}

To illustrate the interpretive value of the DAG-ness framework, we present three representative case studies drawn from the domain-specific outlier lists.

\subsubsection*{Case Study 1: A Hierarchy-Dominated Animal Social Network}
One animal social network exhibits $A = 0.9635$, $F = 0.7418$, $M = 0.0417$, and $S = 0.0699$. This profile indicates a nearly acyclic dominance hierarchy with minimal cyclic locality and limited pathway complexity. The low $M$ suggests that the GSCC is extremely small, consistent with a social system dominated by linear dominance chains rather than reciprocal interactions. The low $S$ further indicates that the network lacks deep cascades or rich pathway structure. This network exemplifies how animal social systems can approach DAG-like structure while retaining small pockets of feedback.

\subsubsection*{Case Study 2: A Feedback-Rich Brain Network}
A brain network with $A = 0.7211$, $F = 0.6526$, $M = 0.0866$, and $S = 0.0141$ represents a feedback-dominated outlier within the brain domain. The low $M$ and $S$ values indicate a small cyclic core and limited pathway complexity, while the moderate $A$ and $F$ values suggest partial hierarchy. This profile is consistent with recurrent microcircuits observed in certain brain regions~\cite{brain-insect2023}. The network's deviation from the domain mean highlights the diversity of structural organization within the brain domain.

\subsubsection*{Case Study 3: A Pathway-Complexity Outlier in Transportation}
A transportation network with $A = 0.9210$, $F = 0.8467$, $M = 0.4139$, and $S = 0.1600$ represents a pathway-complexity outlier. The elevated $S$ indicates rich pathway structure, likely reflecting the presence of multiple alternative routes and redundant connections in the transportation system. The moderate $M$ suggests a nontrivial cyclic core, while the high $A$ and $F$ values indicate substantial hierarchical organization. This network illustrates how engineered systems can exhibit complex pathway structure despite strong design constraints.
\FloatBarrier
\section{Conclusion}
\label{sec:conclusion}

This paper provides the first large-scale empirical evaluation of the refined
4-dimensional DAG-ness framework introduced in \cite{csikos2026dagness}.
By applying the orthogonal components $A(G)$, $F(G)$, $M(G)$, and $S(G)$
to a corpus of 107 directed networks spanning twelve domains, we have
demonstrated that the framework captures meaningful structural variation
across real-world systems while avoiding the topological redundancy and
spectral instabilities present in earlier formulations.

\subsection{Summary of Findings}

Our analysis reveals several consistent patterns across domains:

\begin{itemize}
    \item \textbf{Most real-world networks are nearly acyclic.}
    The acyclicity component $A(G)$ is strongly right-skewed, indicating
    that large-scale directed feedback is rare even in complex systems.

    \item \textbf{Hierarchical flow varies substantially across domains.}
    The flow alignment component $F(G)$ distinguishes domains with
    coherent directional structure (e.g., biological, ecological) from those
    with pervasive bidirectionality (e.g., communication, technological).

    \item \textbf{Macroscopic cyclic traps are domain-specific.}
    The locality component $M(G)$ exhibits a bimodal distribution,
    reflecting the presence or absence of large strongly connected
    components. This behavior validates the Giant Component principle
    developed in \cite{csikos2026dagness}.

    \item \textbf{Dynamical recurrence spans a wide spectrum.}
    The pathway complexity component $S(G)$ captures the diversity of
    spectral behavior across domains, from tree-like systems with
    $\rho(A) \approx 0$ to dense recurrent cores with large spectral radii
    \cite{hornjohnson2013}.
\end{itemize}

These component-level patterns collectively demonstrate that the refined
framework successfully isolates distinct structural and dynamical properties
of directed networks.

\subsection{Cross-Domain Structure and Archetypes}

Beyond marginal behavior, our cross-domain analysis reveals deeper
regularities. Principal component analysis and hierarchical clustering
(Section~\ref{sec:landscape_archetypes}) show that networks from disparate domains
often converge toward similar structural signatures. This convergence is
driven not by domain semantics but by shared topological motifs such as
layered flow, sparse back-edges, or dense recurrent cores.

The structural archetypes identified in Section~\ref{sec:landscape_archetypes}
provide a unifying perspective on these patterns. Despite the diversity of
the dataset, most networks fall into one of four archetypes characterized by
distinct combinations of acyclicity, flow alignment, macroscopic locality,
and spectral recurrence. These archetypes offer a compact vocabulary for
describing directed structure across domains and may serve as a foundation
for future comparative studies.

\subsection{Limitations and Future Directions}

While the refined DAG-ness framework provides a robust and interpretable
measure of directed structure, several limitations remain:

\begin{itemize}
    \item The framework is intentionally structural and does not incorporate
    edge weights, temporal dynamics, or probabilistic interactions.
    Extending the components to weighted or temporal networks is a
    promising direction for future work.

    \item The composite score $D(G)$ uses uniform weights. Although this
    choice provides a neutral baseline, domain-specific or data-driven
    weightings may yield more nuanced insights.

    \item The dataset, while diverse, is not exhaustive. Additional domains
    such as financial transaction networks, neural microcircuits, or
    large-scale social platforms may reveal new structural patterns.
\end{itemize}

Future research may also explore the use of DAG-ness components as
features for machine learning tasks, such as network classification,
anomaly detection, or generative modeling.

\subsection{Synthesis of the DAG-ness Framework}

The empirical results presented in this study serve as the culmination of the DAG-ness framework. The initial development of a continuous measure for directed acyclicity established the necessity of moving beyond binary classification, yet early multi-component iterations were constrained by topological redundancy. By refining the measure into four strictly orthogonal dimensions---explicitly separating the absolute volume of feedback ($A$) from its macroscopic severity ($M$), and topological flow ($F$) from dynamical recurrence ($S$)---the framework achieved the mathematical stability required for large-scale application.

The discovery of the four structural archetypes in this study was only made possible by this rigorous orthogonalization. Under previous, collinear models, the structural boundaries between engineered flow systems (which strictly optimize for throughput) and recurrent biological systems (which rely on modular feedback) would have been blurred by overlapping cyclic penalties. By demonstrating that these diverse generative mechanisms naturally converge into distinct, predictable quadrants of the $(A, F, M, S)$ space, this evaluation validates the theoretical necessity of the orthogonal framework. 

Ultimately, this study proves that DAG-ness is not merely a mathematical abstraction, but a universal, continuous landscape that governs the organization of directed networks across the physical, biological, and informational sciences. By providing a continuous, interpretable, and scalable measure of DAG-likeness, the refined framework offers a powerful tool for analyzing hierarchical structure, diagnosing cyclic traps, and comparing complex systems across domains.

\clearpage
\appendix


\section{Dataset Inventory}
\label{appendix:dataset-inventory}

\begingroup
\small
\setlength{\tabcolsep}{5pt}

\begin{longtable}{%
  >{\raggedright\arraybackslash}p{2.6cm}%
  >{\raggedright\ttfamily\hyphenchar\font=`\-\arraybackslash}p{6.4cm}%
  >{\raggedright\arraybackslash}p{4.0cm}%
  >{\raggedright\arraybackslash}l%
}

\caption[Dataset inventory by domain (107 networks, 12 domains)]%
        {Complete network dataset inventory, grouped by domain.
         Dataset names correspond to filenames used throughout the study.
         \label{tab:appendix-a-datasets}}\\

\toprule
\textbf{Domain} &
\textbf{Dataset (filename)} &
\textbf{Source / Repository} &
\textbf{Citation} \\
\midrule
\endfirsthead

\multicolumn{4}{c}{\tablename\ \thetable\
  \textit{(continued from previous page)}}\\[4pt]
\toprule
\textbf{Domain} &
\textbf{Dataset (filename)} &
\textbf{Source / Repository} &
\textbf{Citation} \\
\midrule
\endhead

\midrule
\multicolumn{4}{r}{\textit{Continued on next page\ldots}}\\
\endfoot

\bottomrule
\endlastfoot

Animal Social
  & aves-songbird-social           & Network Repository & \cite{animal-nr} \\
  & aves-wildbird-network          & Network Repository & \cite{animal-nr} \\
  & insecta-ant-colony2            & Network Repository & \cite{animal-nr} \\
  & insecta-beetle-group-c1-period-2 & Network Repository & \cite{animal-nr} \\
  & mammalia-bat-roosting-indiana  & Network Repository & \cite{animal-nr} \\
  & mammalia-bison-dominance       & Network Repository & \cite{animal-nr} \\
  & mammalia-dolphin-floridatravel & Network Repository & \cite{animal-nr} \\
  & mammalia-hyena-networkb        & Network Repository & \cite{animal-nr} \\
  & mammalia-macaque-dominance     & Network Repository & \cite{animal-nr} \\
  & mammalia-primate-association   & Network Repository & \cite{animal-nr} \\
  & mammalia-raccoon-proximity     & Network Repository & \cite{animal-nr} \\
  & mammalia-voles-bhp-trapping    & Network Repository & \cite{animal-nr} \\
\midrule

Brain
  & bn-fly-drosophila\_medulla\_1  & Winding et al.\ (2023)         & \cite{brain-insect2023}   \\
  & bn-mouse\_brain\_1             & Budapest Connectome (2015)     & \cite{brain-budapest2015} \\
  & bn-mouse\_retina\_1            & Budapest Connectome (2015)     & \cite{brain-budapest2015} \\
  & celegansneural                 & White et al.\ (1986)           & \cite{brain-celegans1986} \\
  & cintestinalis                  & Ryan et al.\ (2016)            & \cite{brain-ciona2016}    \\
  & female\_20k                    & Budapest Connectome (2015)     & \cite{brain-budapest2015} \\
  & fly\_larva                     & Winding et al.\ (2023)         & \cite{brain-insect2023}   \\
  & male\_20k                      & Budapest Connectome (2015)     & \cite{brain-budapest2015} \\
\midrule

Collaboration
  & arXiv                & arXiv / SNAP                   & \cite{snap-leskovec2007}      \\
  & AstroPh              & arXiv / SNAP                   & \cite{snap-leskovec2007}      \\
  & cofe                 & CoFe Dataset                   & \cite{collab-cofe2021}       \\
  & GrQc                 & arXiv / SNAP                   & \cite{snap-leskovec2007}      \\
  & HepPh                & arXiv / SNAP                   & \cite{snap-leskovec2007}      \\
  & HepTh                & arXiv / SNAP                   & \cite{snap-leskovec2007}      \\
  & jazz\_collab         & Gleiser \& Danon (2003)        & \cite{collab-jazz2003}       \\
  & netscience           & Newman (2006)                  & \cite{collab-netscience2006} \\
  & new\_zealand\_collab & Turnbull et al.\ (2018)        & \cite{collab-nz2018}         \\
\midrule

Economic
  & econ-beacxc   & Network Repository & \cite{econ-nr} \\
  & econ-beaflw   & Network Repository & \cite{econ-nr} \\
  & econ-beause   & Network Repository & \cite{econ-nr} \\
  & econ-mbeacxc  & Network Repository & \cite{econ-nr} \\
  & econ-mbeaflw  & Network Repository & \cite{econ-nr} \\
  & econ-mbeause  & Network Repository & \cite{econ-nr} \\
  & econ-orani678 & Network Repository & \cite{econ-nr} \\
  & econ-psmigr1  & Network Repository & \cite{econ-nr} \\
\midrule

Informational
  & darwin                 & Milo et al.\ (2004)      & \cite{info-milo2004}           \\
  & french                 & Milo et al.\ (2004)      & \cite{info-milo2004}           \\
  & google                 & Palla et al.\ (2007)     & \cite{info-palla2007}          \\
  & japanese               & Milo et al.\ (2004)      & \cite{info-milo2004}           \\
  & polblogs               & Adamic \& Glance (2005)  & \cite{info-politicalblogs2005} \\
  & spanish                & Milo et al.\ (2004)      & \cite{info-milo2004}           \\
  & unicodelang            & Unicode Consortium        & \cite{info-unicode2016}        \\
  & webkb\_cornell\_cocite & CMU WebKB (1998)         & \cite{info-webkb1998}          \\
\midrule

Metabolic
  & links\_ath & Huss \& Bhattacharya (2007) & \cite{metabolic-huss2007} \\
  & links\_bja & Huss \& Bhattacharya (2007) & \cite{metabolic-huss2007} \\
  & links\_bme & Huss \& Bhattacharya (2007) & \cite{metabolic-huss2007} \\
  & links\_ecs & Huss \& Bhattacharya (2007) & \cite{metabolic-huss2007} \\
  & links\_hsa & Huss \& Bhattacharya (2007) & \cite{metabolic-huss2007} \\
  & links\_mmu & Huss \& Bhattacharya (2007) & \cite{metabolic-huss2007} \\
  & links\_mtu & Huss \& Bhattacharya (2007) & \cite{metabolic-huss2007} \\
  & links\_pae & Huss \& Bhattacharya (2007) & \cite{metabolic-huss2007} \\
  & links\_rno & Huss \& Bhattacharya (2007) & \cite{metabolic-huss2007} \\
  & links\_sfl & Huss \& Bhattacharya (2007) & \cite{metabolic-huss2007} \\
  & links\_sme & Huss \& Bhattacharya (2007) & \cite{metabolic-huss2007} \\
  & links\_son & Huss \& Bhattacharya (2007) & \cite{metabolic-huss2007} \\
  & links\_stm & Huss \& Bhattacharya (2007) & \cite{metabolic-huss2007} \\
  & links\_sty & Huss \& Bhattacharya (2007) & \cite{metabolic-huss2007} \\
  & links\_ype & Huss \& Bhattacharya (2007) & \cite{metabolic-huss2007} \\
\midrule

Power
  & power-1138-bus  & Network Repository & \cite{power-nr} \\
  & power-494-bus   & Network Repository & \cite{power-nr} \\
  & power-662-bus   & Network Repository & \cite{power-nr} \\
  & power-685-bus   & Network Repository & \cite{power-nr} \\
  & power-bcspwr09  & Network Repository & \cite{power-nr} \\
  & power-bcspwr10  & Network Repository & \cite{power-nr} \\
  & power-eris1176  & Network Repository & \cite{power-nr} \\
  & power-US-Grid   & Network Repository & \cite{power-nr} \\
\midrule

Protein Interactomes
  & 3694  & Zitnik et al.\ (2019) & \cite{protein-zitnik2019} \\
  & 3702  & Zitnik et al.\ (2019) & \cite{protein-zitnik2019} \\
  & 3847  & Zitnik et al.\ (2019) & \cite{protein-zitnik2019} \\
  & 4932  & Zitnik et al.\ (2019) & \cite{protein-zitnik2019} \\
  & 6239  & Zitnik et al.\ (2019) & \cite{protein-zitnik2019} \\
  & 7227  & Zitnik et al.\ (2019) & \cite{protein-zitnik2019} \\
  & 7955  & Zitnik et al.\ (2019) & \cite{protein-zitnik2019} \\
  & 8364  & Zitnik et al.\ (2019) & \cite{protein-zitnik2019} \\
  & 9615  & Zitnik et al.\ (2019) & \cite{protein-zitnik2019} \\
\midrule

Social
  & congress\_network      & Fink et al.\ (2023)  & \cite{social-fink2023} \\
  & copresence-InVS13      & Network Repository   & \cite{nr}              \\
  & copresence-LH10        & Network Repository   & \cite{nr}              \\
  & email-dnc-corecipient  & Network Repository   & \cite{nr}              \\
  & email-enron-large      & Network Repository   & \cite{nr}              \\
  & email-EU               & Network Repository   & \cite{nr}              \\
  & email-univ             & Network Repository   & \cite{nr}              \\
  & soc-epinions           & Network Repository   & \cite{nr}              \\
  & soc-wiki-elec          & Network Repository   & \cite{nr}              \\
 \midrule

Technological
  & 20071112\_caida\_as  & CAIDA                      & \cite{tech-caida}           \\
  & 31\_gnutella         & Ripeanu et al.\ (2002)     & \cite{tech-gnutella2002}    \\
  & internet\_as         & Percolation Dataset (2014) & \cite{tech-percolation2014} \\
  & linux                & KONECT                     & \cite{konect-kunegis2016}   \\
  & python\_dependency   & Python Package Index        & \cite{tech-python}          \\
  & tech-pgp             & Network Repository         & \cite{nr}                   \\
  & tech-routers-rf      & Network Repository         & \cite{nr}                   \\
  & tech-WHOIS           & Network Repository         & \cite{nr}                   \\
  \midrule

Transportation
  & openflights\_2010              & Opsahl (2011)               & \cite{transport-opsahl2011}   \\
  & us\_airport\_network\_top\_500 & Colizza et al.\ (2007)      & \cite{transport-colizza2007}  \\
  & euroroad                       & \v{S}ubelj \& Bajec (2011) & \cite{transport-subelj2011}   \\
  & eu\_airlines                   & Cardillo et al.\ (2013)     & \cite{transport-cardillo2013} \\
  & faa\_routes                    & FAA (2010)                  & \cite{transport-faa2010}      \\
  & paris\_transportation          & Asgari et al.\ (2016)       & \cite{transport-asgari2016}   \\
  & road-belgium-osm               & \v{S}ubelj \& Bajec (2011) & \cite{transport-subelj2011}   \\
  & road-luxembourg-osm            & \v{S}ubelj \& Bajec (2011) & \cite{transport-subelj2011}   \\
\midrule

Water Distribution
  & Anytown                & Walski et al.\ (2016)  & \cite{water-walski2016}   \\
  & CalibrationNetworks    & Ostfeld et al.\ (2016) & \cite{water-ostfeld2016}  \\
  & LongTermImprovement    & Marchi et al.\ (2021)  & \cite{water-marchi2021}   \\
  & Water Sensor Network 1 & Ostfeld et al.\ (2021) & \cite{water-ostfeld2021a} \\
  & Water Sensor Network 2 & Ostfeld et al.\ (2021) & \cite{water-ostfeld2021b} \\

\end{longtable}
\endgroup

\clearpage
\section{Domain-Specific Outliers}
\label{app:domain-outliers}

\begin{longtable}{lccccc}
\caption{Top five domain-specific outliers for each domain. Outlier scores are computed using domain-standardized $z$-scores across the four DAG-ness components $(A, F, M, S)$.} \\
\label{tab:domain-outliers}
\\
\toprule
\textbf{Domain} & \textbf{A} & \textbf{F} & \textbf{M} & \textbf{S} & \textbf{Outlier Score} \\
\midrule
\endfirsthead

\toprule
\textbf{Domain} & \textbf{A} & \textbf{F} & \textbf{M} & \textbf{S} & \textbf{Outlier Score} \\
\midrule
\endhead

\midrule
\multicolumn{6}{r}{\textit{Continued on next page}} \\
\endfoot

\bottomrule
\endlastfoot

Animal Social Networks & 0.9635 & 0.7418 & 0.0417 & 0.0699 & 2.1645 \\
Animal Social Networks & 0.9796 & 0.7016 & 0.7391 & 0.0734 & 2.0229 \\
Animal Social Networks & 0.9704 & 0.9307 & 0.2338 & 0.0707 & 1.6240 \\
Animal Social Networks & 0.9929 & 0.7737 & 0.2538 & 0.0151 & 1.4556 \\
Animal Social Networks & 1.0000 & 1.0000 & 1.0000 & 1.0000 & 0.9632 \\

Brain Networks & 1.0000 & 1.0000 & 1.0000 & 1.0000 & 2.2890 \\
Brain Networks & 0.7211 & 0.6526 & 0.0866 & 0.0141 & 1.4718 \\
Brain Networks & 0.7630 & 0.6234 & 0.0000 & 0.0096 & 1.2483 \\
Brain Networks & 0.7853 & 0.7096 & 0.0294 & 0.0483 & 1.1779 \\
Brain Networks & 1.0000 & 0.7401 & 1.0000 & 0.5000 & 1.1437 \\

Collaboration Networks & 0.8863 & 0.8517 & 0.7755 & 0.0247 & 2.8284 \\
Collaboration Networks & 1.0000 & 1.0000 & 1.0000 & 1.0000 & 1.1179 \\
Collaboration Networks & 1.0000 & 1.0000 & 1.0000 & 1.0000 & 1.1179 \\
Collaboration Networks & 1.0000 & 1.0000 & 1.0000 & 1.0000 & 1.1179 \\
Collaboration Networks & 1.0000 & 1.0000 & 1.0000 & 1.0000 & 1.1179 \\

Economic Networks & 0.9759 & 0.9422 & 0.0004 & 0.4414 & 2.2789 \\
Economic Networks & 0.9221 & 0.7881 & 0.1792 & 0.7115 & 1.3971 \\
Economic Networks & 0.9209 & 0.7860 & 0.1708 & 0.7115 & 1.3971 \\
Economic Networks & 0.9507 & 0.8537 & 0.2490 & 0.0000 & 1.0564 \\
Economic Networks & 0.9209 & 0.7860 & 0.1708 & 0.0001 & 0.8676 \\

Informational Networks & 1.0000 & 0.5000 & 1.0000 & 0.0074 & 2.1503 \\
Informational Networks & 1.0000 & 1.0000 & 1.0000 & 1.0000 & 1.8837 \\
Informational Networks & 1.0000 & 1.0000 & 1.0000 & 1.0000 & 1.8837 \\
Informational Networks & 0.7389 & 0.7160 & 0.1549 & 0.0164 & 1.5067 \\
Informational Networks & 0.7703 & 0.7424 & 0.3520 & 0.0179 & 1.2207 \\

Metabolic Networks & 0.9533 & 0.9024 & 0.7004 & 0.0631 & 1.9769 \\
Metabolic Networks & 0.9547 & 0.9023 & 0.7073 & 0.0638 & 1.9587 \\
Metabolic Networks & 0.9689 & 0.9267 & 0.7692 & 0.0794 & 1.5058 \\
Metabolic Networks & 0.9574 & 0.9131 & 0.7679 & 0.0769 & 1.4440 \\
Metabolic Networks & 0.9709 & 0.9218 & 0.7528 & 0.0799 & 1.4084 \\

Power Grids & 1.0000 & 0.5000 & 1.0000 & 0.1675 & 1.5010 \\
Power Grids & 1.0000 & 0.5000 & 1.0000 & 0.1467 & 1.2225 \\
Power Grids & 1.0000 & 0.5000 & 1.0000 & 0.1179 & 0.8353 \\
Power Grids & 1.0000 & 0.5000 & 1.0000 & 0.0001 & 0.7463 \\
Power Grids & 1.0000 & 0.5000 & 1.0000 & 0.0001 & 0.7463 \\

Protein Interactomes & 1.0000 & 1.0000 & 1.0000 & 1.0000 & NaN \\
Protein Interactomes & 1.0000 & 1.0000 & 1.0000 & 1.0000 & NaN \\
Protein Interactomes & 1.0000 & 1.0000 & 1.0000 & 1.0000 & NaN \\
Protein Interactomes & 1.0000 & 1.0000 & 1.0000 & 1.0000 & NaN \\
Protein Interactomes & 1.0000 & 1.0000 & 1.0000 & 1.0000 & NaN \\

Social Networks & 0.8361 & 0.6770 & 0.0274 & 0.0276 & 2.2544 \\
Social Networks & 1.0000 & 0.5000 & 1.0000 & 0.0000 & 1.8764 \\
Social Networks & 0.8747 & 0.7640 & 0.6020 & 0.0192 & 1.2438 \\
Social Networks & 1.0000 & 1.0000 & 1.0000 & 1.0000 & 1.2074 \\
Social Networks & 1.0000 & 1.0000 & 1.0000 & 1.0000 & 1.2074 \\

Technological Networks & 0.9532 & 0.9532 & 0.7739 & 0.0000 & 2.8436 \\
Technological Networks & 1.0000 & 1.0000 & 1.0000 & 1.0000 & 2.7706 \\
Technological Networks & 1.0000 & 0.5000 & 1.0000 & 0.0000 & 1.1603 \\
Technological Networks & 1.0000 & 0.5000 & 1.0000 & 0.0349 & 1.1603 \\
Technological Networks & 1.0000 & 0.5000 & 1.0000 & 0.0066 & 1.1603 \\

Transportation Networks & 0.9210 & 0.8467 & 0.4139 & 0.1600 & 2.2654 \\
Transportation Networks & 1.0000 & 1.0000 & 1.0000 & 1.0000 & 1.4536 \\
Transportation Networks & 1.0000 & 1.0000 & 1.0000 & 1.0000 & 1.4536 \\
Transportation Networks & 0.9285 & 0.5102 & 0.9728 & 0.0158 & 1.3598 \\
Transportation Networks & 1.0000 & 0.5000 & 1.0000 & 0.0204 & 0.8840 \\

Water Distribution Networks & 1.0000 & 1.0000 & 1.0000 & 1.0000 & 1.7307 \\
Water Distribution Networks & 0.9660 & 0.9068 & 0.8672 & 0.4333 & 1.7162 \\
Water Distribution Networks & 0.9860 & 0.9860 & 0.9823 & 0.4562 & 0.5449 \\
Water Distribution Networks & 0.9865 & 0.9865 & 0.9828 & 0.4562 & 0.5449 \\

\end{longtable}

\end{document}